\documentclass[12pt]{iopart}

\usepackage{epsfig}
\usepackage{graphicx}
\usepackage{dcolumn}
\usepackage{bm}

\def\gapp{\lower.35em\hbox{$\stackrel{\textstyle>}{\sim}$}}
\def\lapp{\lower.35em\hbox{$\stackrel{\textstyle<}{\sim}$}}

\begin{document}
\title{Effect of Holstein phonons on the electronic properties
of graphene}
\author{T.~Stauber$^{1,2}$ and  N.~M.~R. Peres$^2$}

\address{$^1$Instituto de Ciencia de Materiales de
Madrid. CSIC. Cantoblanco. E-28049 Madrid, Spain}

\address{$^2$Center of Physics and Department of
Physics, University of Minho, P-4710-057, Braga, Portugal}


\date{\today}
\begin{abstract}
We obtain the self-energy of the electronic propagator due to the
presence of Holstein polarons within the first Born
approximation. This leads to a renormalization of the Fermi velocity
of one percent. We further compute the optical conductivity of the
system at the Dirac point and at finite doping within the
Kubo-formula. We argue that the effects due to Holstein phonons are
negligible and that the Boltzmann approach which does not include
inter-band transition and can thus not treat optical phonons due to
their high energy of $\hbar\omega_0\sim0.1-0.2$eV, remains valid.
\end{abstract}
%
\submitto{\JPCM}
%
%
%

\section{Introduction}
Three years ago, Geim, Novoselov and co-workers have succeeded in
isolating and contacting a single layer of graphite
(graphene).\cite{Nov04} Contrary to common wisdom, this experiment
showed that true two-dimensional lattices are thermodynamically
stable.  \cite{Nov07,Stein07} This stability comes about because the
system gently crumples to the third direction forming ripples.
\cite{Jannik07} It is therefore an important problem to study the
effect of out-plane phonons on the electronic properties of the
system.  Contrary to two-dimensional electron-systems in semiconductor
heterostructures, where (scalar) electrons interact with
bulk\cite{2Dpolarons} or surface\cite{ConfinedPhonons} phonons, in
graphene there are two-dimensional in-plane as well as out-of-plane
vibrational modes to which Dirac (spinor) fermions will couple.

There are two different vertex-types modeling the interaction 
of electrons with optical phonons. First, due to the atomic 
displacement within the plane, the tunneling-matrix element 
between two carbon atoms varies. This gives rise to a 
Su-Schrieffer-Heeger type coupling (current-current coupling).\cite{Su79} 
The two-dimensional gauge field is composed by the longitudinal 
optical (LO) and transverse optical (TO) phonons which are 
degenerate at $q=0$. This vertex type is also found from symmetry 
arguments.\cite{Manes07} There are also out-of-plane vibrations. 
These lattice displacements are symmetric with respect to their 
neighboring atoms and thus couple to the electronic densities. 
The optical (ZO) modes can then be described within the Holstein 
model.\cite{Holstein59} For a recent account on the Green's function of the Holstein polaron, see Ref. \cite{Goodvin06} and references therein.

Out-of-plane modes are energetically 
smaller than in-plane vibrations due to the hybridization of 
the $sp^2$-orbitals within the graphene sheet. A tight-binding 
calculation including nearest and second-nearest neighbors 
yields $\omega_{\rm ZO}\approx\omega_{\rm LO}/2$ for the 
optical branch close to the $\Gamma$-point.\cite{Fal07} 
For a first-principle calculation of the phonon-spectra, 
see the work by Wirtz and 
Rubio.\cite{rubio} The effect of the electron-phonon coupling
on the local density of states of zig-zag graphene ribbons has
been studied by Sasaki {\it et al.}.\cite{Sasaki07}
The effect of the optical phonons on the Raman spectrum of
disordered graphene was studied by Castro Neto and Guinea.
\cite{Castro07}

The effect of Holstein phonons on transport properties has been
receiving renewed interest in the context of the one-dimensional
Holstein-Hubbard model.\cite{Fehske07} Here, we will discuss Holstein
phonons in graphene for the following reason. It is currently believed
that transport properties can be well described within a
semi-classical Boltzmann approach.\cite{Nomura06,Adam07,Peres07} This
implies well-defined quasi-particles, i.e., only one band is
considered and inter-band transitions are ruled out. Doing so, several
scattering mechanisms have been discussed, ranging from local defects
(vacancies or substitutions), long-ranged Coulomb impurities in the
substrate or due to adsorbed atoms to acoustical
phonons.\cite{Sta07,Hwang06}
 
Optical phonons cannot be treated within the one-band Boltzmann
approach\cite{AK07} since they would induce interband-transitions at typical
densities of $n\lapp 5\times10^{12}$cm${}^{-2}$. It is therefore
crucial to assess this scattering mechanism via the Kubo-formalism. In
section II, we present our model for Holstein phonons and calculate
the electronic self-energy within the first Born approximation in
section III. We further compute the optical conductivity in section
IV, using the full Greens function, but neglecting vertex
corrections. We close with conclusions and outlook.


\section{The effective model}
\label{sec_model}
In two-dimensional graphene sheets, first-principle calculations 
reveal that the Born-Oppenheimer approximation is not valid for 
doped graphene sheets.\cite{Laz06} Fr\"ohlich polarons\cite{Fro54} 
are thus not a good starting point to describe electron-lattice 
interaction in graphene. Here, we present the study of the 
electron-lattice coupling due to localized Holstein-phonons 
in a two-dimensional honeycomb lattice, thus treating the 
ZO-phonons (out-of-plane vibrations).

The honeycomb lattice (the lattice of graphene) 
is made of two interpenetrating triangular
lattices, defining two non-equivalent sites, usually labeled as
$A$ and $B$ sites.\cite{Peres06} 
The model Hamiltonian for ZO-polarons in graphene reads as follows:
\begin{equation}
H-t\sum_{i,\sigma \bm \delta}
\left[
a^\dag_\sigma(\bm R_i)b_\sigma(\bm R_i+\bm \delta) + {\rm H. c.}
\right]+
\sum_{\bm q}\omega_{\bm q}c^\dag_{\bm q}c_{\bm q}
+ V_{ep}
\label{Hamiltonian}
\end{equation}
where $a^\dag_\sigma(\bm R_i)$ ($b^\dag_\sigma(\bm R_i)$) creates
an electron at an atom of the $A$(B) sub-lattice and $c^\dag_{\bm q}$
are creation phonon operators.  
The energy $\omega_{\bm q}$ is the dispersion of the ZO-phonon and $V_{ep}$
the electron-phonon interaction. We model the coupling to the ZO-phonons as the usually density-density coupling\cite{Mahan}
\begin{equation}
V_{ep}=D\sum_{\sigma,i,j}
\left[
a^\dag_\sigma(\bm R_i)a_\sigma(\bm R_i)+
b^\dag_\sigma(\bm R_i)b_\sigma(\bm R_i)
-1\right]Q_j\,,
\end{equation}
with $Q_j$ defined as 
\begin{equation}
Q_j=\sum_{\bm q}X_{\bm q}e^{i\bm q \cdot \bm R_j}
(c_{\bm q}+c^\dag_{-\bm q})\,,
\end{equation}
and 
\begin{equation}
X_{\bm q} = \sqrt{\frac {\hbar^2} {2MN\omega_{\bm q}} }
\end{equation}
and $M$ the ion's mass and $N$ the number of unit cells in the crystal. Note that we couple the density with respect to the half-filled band such that particle-hole symmetry is conserved by reversing the sign of the coupling constant. In the following, though, we shall neglect this shift in the bosonic operators.

The effect of  Holstein polarons is obtained using 
$\omega_{\bm q}\simeq \omega_0$ and transforming the operators
in Hamiltonian (\ref{Hamiltonian}) to momentum space. This gives
\begin{equation}
H =-t\sum_{\bm q,\sigma}
\left(
\phi_{\bm q}a^\dag_{\bm q,\sigma}
b_{\bm q,\sigma} + {\rm H. c.}
\right)
+\sum_{\bm q}\omega_0c^\dag_{\bm q}c_{\bm q}
+V_{ep}\,,
\end{equation}
with $\phi_{\bm q} =\sum_{\bm \delta}e^{-i\bm \delta\cdot\bm q}$, 
$\bm \delta$ the vectors connecting the three nearest neighbors
on the honeycomb lattice,
and $V_{ep}$ given by
\begin{equation}
V_{ep} =D\sum_{\bm p,\bm q,\sigma}
X_{\bm q}
\left(
a^\dag_{\bm p,\sigma}a_{\bm p+\bm q,\sigma}
+
b^\dag_{\bm p,\sigma}b_{\bm p+\bm q,\sigma}\right)
\left(
c_{\bm q}+c^\dag_{-\bm q}
\right)\;.
\label{PhononCoupling}
\end{equation}

Let us comment on the coupling constant $D$. Due to the mirror
symmetry of the graphene-sheet, one might think that the linear
coupling to lattice displacements is zero. But the mirror symmetry is
broken for samples where graphene lies on top of a SiO${}_2$- or
SiC-substrate (for a discussion, see Ref. \cite{Manes07}). To quantify
the coupling constant in terms of the dimensionless constant
$g=\sqrt N DX_0/\omega_0$, we assume that the coupling mechanism is
due to a variation of the hopping matrix element. This yields $g$ of
the order of unity.\cite{Castro07}


\section{Second order perturbation theory}
\label{sec_perturbation}
If the phonon energy scale is much smaller than the electronic energy
scale, Migdal's theorem states that it is sufficient to calculate the
lowest-order self-energy diagram.\cite{Migdal58} This diagram can
further be calculated using the bare electron Greens function. Still,
the importance of electron-phonon coupling also depends on the
dimensionality of the system. E.g., in self-assembled quantum dots,
vertex ''corrections'' to the polarization of an electron-hole pair
give rise to charge-cancellation,\cite{SchmittRink87} thus changing
the optical conductivity significantly.\cite{Sta06} Also for $A_3{\rm
C}_{60}$ ($A$=K,Rb)\cite{Gun94} and for general one-dimensional
systems,\cite{Med94} Migdal's theorem is not valid. Furthermore,
electron-electron interaction can affect the effective electron-phonon
interaction.\cite{Fehske07} 

In graphene sheets, electron-electron interaction is generally
neglected, i.e., one assumes a ``normal'' ground-state at zero doping
(one electron per unit cell) - characterized by a semi-metal. Since
the average kinetic and interaction energy per particle both scale
with $\sqrt{n}$ where $n$ is the carrier density, the
interaction does not become important at finite doping,
either. Electron-electron interaction is also neglected in recent
works on localization\cite{Aleiner06} even though disorder enhances
the effect of interaction.\cite{Sta05} The same should hold for the
electron-phonon vertex corrections which results in an effective
electron-electron interaction and we thus believe that the assumption
of Migdal's theory is a good starting point to discuss the
Holstein-phonons in graphene.

Because of
the existence of two sub-lattices, the Green's function
needs to be written as a $2\times 2$ matrix:  
\begin{eqnarray}
\bm G_{\sigma}(\bm k,\tau) = \left(\begin{array}{cc}
G_{AA,\sigma} (\bm k, \tau) \hspace{0.5cm} & G_{AB,\sigma}(\bm k,\tau) \\
G_{BA,\sigma} (\bm k,\tau) \hspace{0.5cm} & G_{BB,\sigma}(\bm k,\tau)  
\end{array}\right) \, ,
\end{eqnarray}
with
\begin{eqnarray}
G_{AA,\sigma}(\bm k, \tau) &=& - \langle {\cal T} a_{\bm k,\sigma}(\tau) 
a_{\bm k,\sigma}^{\dag}(0)
\rangle \, ,
\nonumber
\\
G_{AB,\sigma}(\bm k, \tau) &=& - \langle {\cal T} a_{\bm k,\sigma}(\tau) 
b_{\bm k,\sigma}^{\dag}(0)
\rangle \, ,
\nonumber
\\
G_{BA,\sigma}(\bm k,\tau) &=& - \langle {\cal T} b_{\bm k,\sigma}(\tau) 
a_{\bm k,\sigma}^{\dag}(0)
\rangle \, ,
\nonumber
\\
G_{BB,\sigma}(\bm k,\tau) &=& - \langle {\cal T} b_{\bm k,\sigma}(\tau) 
b_{\bm k,\sigma}^{\dag}(0)
\rangle \, ,
\label{defg}
\end{eqnarray}
where $\tau$ is the ``imaginary'' time, and ${\cal T}$ is the time ordering
operator. 

Up to second order in perturbation theory, and after
transforming the time dependence of the Green's
function to Matsubara frequencies, we obtain
the following result:
\begin{equation}
{\bm G} = {\bm G^{(0)}}+ {\bm G^{(0)}}{\bm\Sigma}{\bm G^{(0)}}\,,
\end{equation}
where the matrix ${\bm\Sigma}$ is defined as
\begin{equation}
\bm \Sigma=\left(
\begin{array}{cc}
 \Sigma_{AA}& \Sigma_{AB}\\
\Sigma_{BA}&\Sigma_{BB}
\end{array}
\right)\,,
\end{equation}
and the matrix element $\Sigma_{\alpha\beta}$ is defined by 
($\alpha,\beta=A,B$)
\begin{equation}
\label{label:selfenergy}
\Sigma_{\alpha\beta}(i\omega_n,\bm p)=-\frac 1 
{\beta}\sum_{\bm q,\nu}D^2X^2_{\bm q}
D^{(0)}(\bm q,i\nu)
 G^{(0)}_{\alpha\beta}(\bm p -\bm q,i\omega_n-i\nu).
\end{equation}

Because both $X_{\bm q}$ and $D^{(0)}(\bm q,i\nu)$ are momentum
independent, i.e., we set $\omega_{\bm q}=\omega_0$
and the phonon propagator  $D^{(0)}(\bm q,i\nu)$ is given by
\begin{equation}
D^{(0)}(\bm q,i\nu) =\frac {2\omega_0}{(i\nu)^2-\omega^2_0}\,,
\end{equation}
the matrix
elements of the self energy matrix can be written in a simplified 
form, reading
$
\Sigma_{\alpha\beta}(i\omega_n)=-\frac 1 {\beta}\sum_{\bm q,\nu}D^2X^2_0
D^{(0)}(i\nu)G^{(0)}_{\alpha\beta}(\bm q,i\omega_n+i\nu)\;
\label{self_energy}
$
where $X_0$ is $X_{\bm q}$ with $\omega_{\bm q}$ replaced by $\omega_0$. Notice that we neglected the constant shift in the self-energy due to the shift of the bosonic displacement operator, still present in Eq. (\ref{label:selfenergy}). This is consistent since we also neglected the Hartree correction to the self energy.

From the above we can write down a Dyson equation for the electronic
propagator, given by
\begin{equation}
{\bm G} = {\bm G^{(0)}}+ {\bm G^{(0)}}{\bm\Sigma}{\bm G}\,,
\label{dyson}
\end{equation}
which has to be solved for ${\bm G}$. The equation giving the Matsubara
Green's function for the free electronic system reads
\begin{equation}
\left(
\begin{array}{cc}
 i\omega_n& -t\phi(\bm k)\\
-t\phi^\ast(\bm k) &i\omega_n
\end{array}
\right){\bm G^{(0)}(i\omega_n,\bm k)}={\bm 1}\,,
\end{equation}
and ${\bm 1}$ is the $2\times2$ unit matrix. 
Within the Dirac cone approximation, which applies in a range
of 1 eV, one has  
$\Sigma_{AB}(i\omega_n)=\Sigma_{BA}(i\omega_n)=0$. Also 
the following result holds 
$\Sigma_{AA}(i\omega_n)=\Sigma_{BB}(i\omega_n)=\Sigma(i\omega_n)$ leading to
a simplified form for the Dyson equation (\ref{dyson}), which can
be readily solved, giving
\begin{equation}
{\bm G} = \frac {
\left(
\begin{array}{cc}
 i\omega_n-\Sigma(i\omega_n)& t\phi(\bm k)\\
t\phi^\ast(\bm k) &i\omega_n-\Sigma(i\omega_n)
\end{array}
\right)
}
{(i\omega_n-\Sigma)(i\omega_n-\Sigma)-t^2\vert \phi\vert^2}\;.
\label{dyson_solved}
\end{equation}
The matrix elements of Eq. (\ref{dyson_solved}) can be put into
a simpler form reading
\begin{eqnarray}
{ G}_{AA}(\omega_n,{\bm k})&=& \sum_{j=\pm 1}
\frac{1/2}{i\omega_n-\Sigma(i\omega_n) - j t\vert \phi(\bm k)\vert}
\label{gaa}\,,\\
{ G}_{AB}(\omega_n,{\bm k})&=& \sum_{j=\pm 1}
\frac{je^{i\delta(\bm k)}/2}
{i\omega_n-\Sigma(i\omega_n) -  j t\vert \phi(\bm k)\vert}\label{gab}\,,\\
{ G}_{BA}(\omega_n,\bm k)&=& \sum_{j=\pm 1}
\frac{je^{-i\delta(\bm k)}/2}
{i\omega_n-\Sigma(i\omega_n) -  j t\vert \phi(\bm k)\vert}\label{gba}\,,\\
{ G}_{BB}(\omega_n,{\bm k})
&=& { G}_{AA}(\omega_n,\bm k)\,.
\label{gbb}
\end{eqnarray}

The summation over the bosonic frequency $\nu$ in Eq. (\ref{self_energy})
can be performed using standard methods leading to
\begin{eqnarray}
\Sigma(i\omega_n)=\sum_{\bm q,j=\pm 1}D^2X^2_0\frac 1 2
\left(
\frac {N_0+n_F(jt\vert \phi(\bm q)\vert)}
{i\omega_n-jt\vert \phi(\bm q)\vert+\omega_0}
+
\frac {N_0+1-n_F(jt\vert \phi(\bm q)\vert)}
{i\omega_n-jt\vert \phi(\bm q)\vert-\omega_0}
\right)\,.
\label{Self_2}
\end{eqnarray}
The integrals over the momentum variable in Eq. (\ref{Self_2})
can be easily computed at zero temperature and  
within the Dirac cone approximation,
yielding an explicit form for the self energy. 
\subsection{Zero doping}
At zero temperature and zero chemical potential (that is at the
neutrality point) the self energy, denoted by
$\Sigma\rightarrow\Sigma_0$, has a simplified form reading
\begin{eqnarray}
\Sigma_0(i\omega_n)=\frac {g^2\omega^2_0}{N\hbar^2}
\sum_{\bm q}\frac 1 2
\left(
\frac 1 {i\omega_n-t\vert\phi(\bm q)\vert/\hbar-\omega_0/\hbar }
+
\frac 1 {i\omega_n+t\vert\phi(\bm q)\vert/\hbar+\omega_0/\hbar }
\right)
\label{SE_ZT}
\end{eqnarray}
where we have introduced the missing $\hbar$'s omitted in the
beginning of this section and $g$ denotes a dimensionless coupling
constant of order unity, defined below
Eq. (\ref{PhononCoupling}). Performing the analytical continuation
$i\omega_n\rightarrow \omega + i\delta$ and computing the momentum
integral in Eq. (\ref{SE_ZT}) one obtains the retarded self energy,
$\Sigma^{\rm ret}(\omega)$, of the polaron problem, which reads
\begin{eqnarray}
&&\Sigma_0^{\rm ret}(\omega)=
\frac {A_c}{2\pi}
\left(
\frac {g\omega_0}{\hbar}
\right)^2
\left(
-\frac {\omega}{v^2_F}\ln
\left\vert
\frac {(v_Fk_c)^2}{\omega^2-(\omega_0/\hbar)^2}
\right\vert
\right.
\nonumber\\
&+&
\left.
\frac {\omega_0/\hbar}{v_F^2}
\ln
\left\vert
\frac {\omega+\omega_0/\hbar}{\omega-\omega_0/\hbar}
\right\vert
\right)
-i\pi
\frac {A_c}{2\pi}
\left(
\frac {g\omega_0}{\hbar}
\right)^2\label{self_energy_ep}\\
&\times&\left[\left(
\frac{\omega}{v^2_F}-\frac {\omega_0}{v^2_F\hbar}
\right)\theta(\omega-\omega_0/\hbar)
\theta(v_Fk_c+\omega_0/\hbar-\omega)
\right.\nonumber\\
&-&\left.
\left(
\frac{\omega}{v^2_F}+\frac {\omega_0}{v^2_F\hbar}
\right)\theta(-\omega-\omega_0/\hbar)
\theta(v_Fk_c+\omega_0/\hbar+\omega)
\right]\,,
\nonumber
\end{eqnarray}
with $\pi k_c^2=(2\pi)^2/A_c$, $A_c=a^23\sqrt 3 /2$, and 
$a=1.42$ \AA.

One can easily obtain the renormalization of the electronic
spectrum due to the retarded self-energy induced by the phonons,
using Rayleigh-Schr\"odinger perturbation theory.\cite{Mahan} 
In this scheme, the $\omega$ dependence of the self-energy is
replace by the bare electronic dispersion. Close to the 
Dirac point, the dispersion is simply given by
$\omega=\pm v_Fk$ where the upper (lower) sign holds for electrons (holes) and we obtain the new energy spectrum
$E(\bm k)$ from
\begin{equation}
E(\bm k)=v_F\hbar k +\hbar\Re \Sigma_0^{\rm ret}(v_F k)\,
\end{equation}
for the conduction band and
\begin{equation}
E(\bm k)=-v_F\hbar k +\hbar\Re \Sigma_0^{\rm ret}(-v_F k)\,
\end{equation}
for the valence band. Electron-hole symmetry is thus preserved, if $\Re \Sigma_0^{\rm ret}(x)=-\Re \Sigma_0^{\rm ret}(-x)$ which is the case (see Eq. (\ref{self_energy_ep})).

\begin{figure}
\begin{center}\includegraphics*[width=8cm]{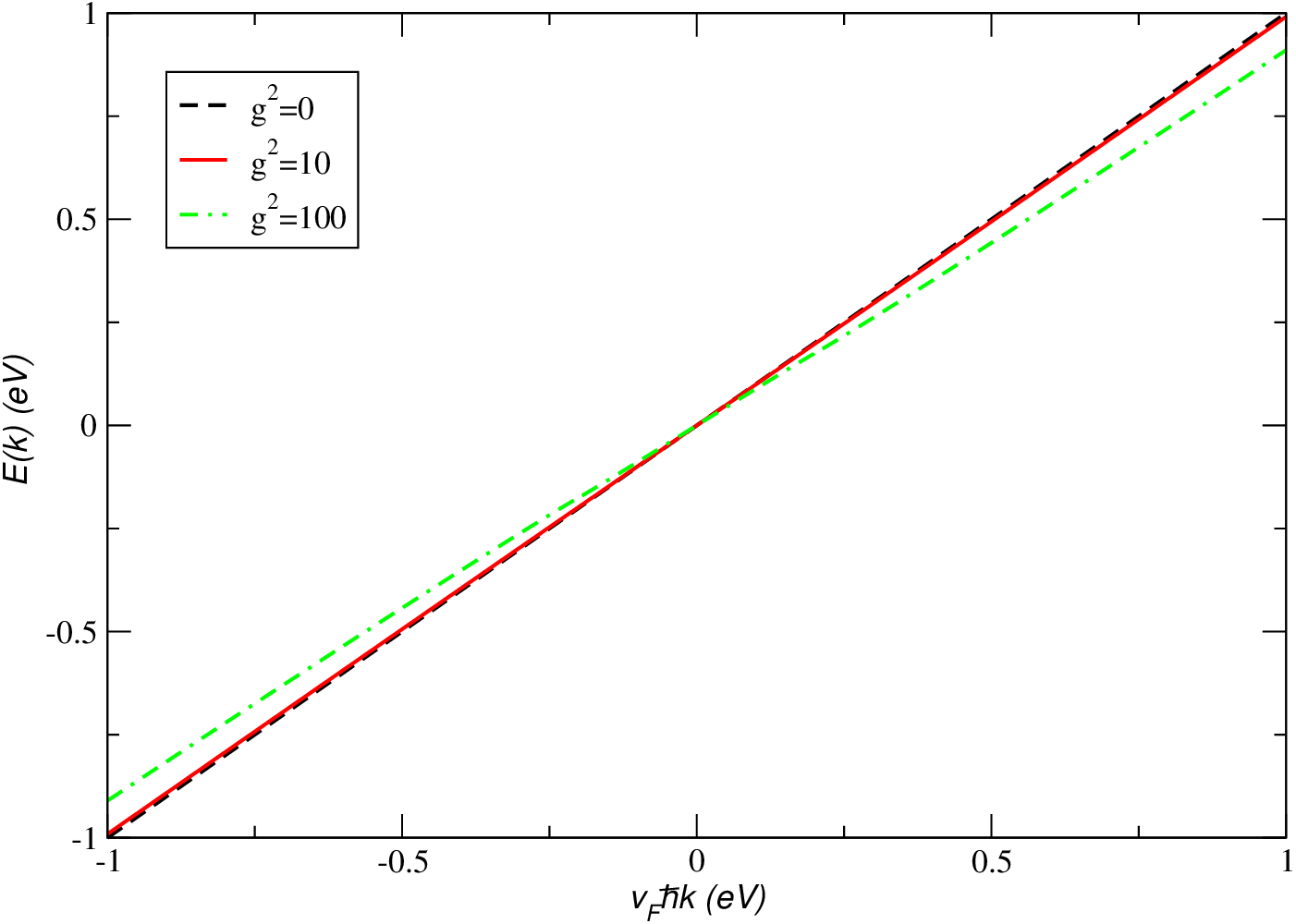}\end{center}
\caption{\label{cap:bilayer}(Color online) Renormalized
energy band at the neutrality point, given by
$E(\bm k)=\pm v_F\hbar k \pm\hbar\Re\Sigma_0^{\rm ret}(v_F k)$ 
for two different values
of the coupling constant $g$.}
\end{figure}

Considering that $v_Fk$ is the smallest
energy in the problem, that is $k$ is very close to the
Dirac point, we obtain for $\Re \Sigma_0^{\rm ret}(v_F k)$
the result
\begin{equation}
\Re \Sigma_0^{\rm ret}(v_F k)=
-v_F k \frac {A_c}{\pi v_F^2}
\left(
\frac {g\omega_0}{\hbar}
\right)^2
\ln \frac{v_Fk_c}{\omega_0/\hbar}\,.
\end{equation} 
Considering the out-of-plane optical mode of graphene\cite{rubio},
which has an energy of $\omega_{ZO}\simeq 0.1$ eV and
using $g^2=10$ we obtain
\begin{equation}
\frac {A_c}{\pi v_F^2}
\left(
\frac {g\omega_0}{\hbar}
\right)^2
\ln \frac{v_Fk_c}{\omega_0/\hbar}\simeq 0.02\;.
\end{equation}
In Fig. \ref{cap:bilayer}, the renormalized energy 
dispersion is shown for various values of the coupling constant. We also included a large coupling constant $g=10$ for better illustration of the effect of the electron-phonon interaction. 
\subsection{Finite doping}
At finite doping, the one-particle dispersion $j\vert\phi(\bm k)\vert$ must be replaced by $j\vert\phi(\bm k)\vert-\mu$ in Eqs. (\ref{gaa}) - (\ref{gbb}) and Eq. (\ref{Self_2}), where $\mu=\hbar v_F k_F$ denotes the Fermi energy. The retarded self-energy can then be written as
$
\Sigma_\mu^{\rm ret}(\omega)=\Sigma_0^{\rm ret}(\omega+\mu/\hbar)+\Delta\Sigma(\omega+\mu/\hbar)
$
where we have
\begin{eqnarray}
\Delta\Sigma(\omega)=\frac {A_c}{2\pi}
\left(
\frac {g\omega_0}{\hbar}
\right)^2
\left(
-\frac{\omega_0/\hbar}{v_F^2}\ln\left|\frac{(\mu/\hbar-\omega)^2-(\omega_0/\hbar)^2}{\omega^2-(\omega_0/\hbar)^2}\right|
\right.
\nonumber\\
\left.
-\frac{\omega}{v_F^2}\ln\left|\frac{(\mu/\hbar-\omega-\omega_0/\hbar)(\omega-\omega_0/\hbar)}{(\mu/\hbar-\omega+\omega_0/\hbar)(\omega+\omega_0/\hbar)}\right|\right)-
\nonumber\\
i\pi
\frac {A_c}{2\pi}
\left(
\frac {g\omega_0}{\hbar}
\right)^2
\left[\left(
\frac{\omega}{v^2_F}+\frac {\omega_0}{v^2_F\hbar}
\right)\theta(\omega+\omega_0/\hbar)
\theta(\mu/\hbar-\omega_0/\hbar-\omega)
\right.
\nonumber\\
\left.
-
\left(
\frac{\omega}{v^2_F}-\frac {\omega_0}{v^2_F\hbar}
\right)\theta(\omega-\omega_0/\hbar)
\theta(\mu/\hbar+\omega_0/\hbar-\omega)
\right]\,.
\end{eqnarray}
We note that for finite doping $\Sigma_\mu^{\rm ret}(\omega)$ diverges logarithmically at $\omega=\pm\omega_0/\hbar$.


\section{Optical conductivity }
\label{sec_optical}
In this section we want to compute the optical conductivity of graphene
and study how the conductivity of the system is affected by the 
out-of-plane (ZO)
phonons. To determine the conductivity one needs to know the current
operator  $j_x$, which is composed of the paramagnetic and 
diamagnetic contributions
$j_x = j_x^P+A_x(t)j_x^D$, each of them given by\cite{Per07}
\begin{eqnarray}
j^P_x = -\frac {itea}{2\hbar}\sum_{\bm k,\sigma}
\left[(\phi(\bm k)-3)a_{\sigma}^{\dagger}(\bm k)b_{\sigma}(\bm k)
 - 
(\phi^\ast(\bm k)-3)b^\dag_{\sigma}(\bm k)
a_{\sigma}(\bm k)
\right]\,,
\end{eqnarray}
and 
\begin{eqnarray}
j^D_x = -\frac {te^2a^2}{4\hbar^2}\sum_{\bm k,\sigma}
\left[(\phi(\bm k)+3)a_{\sigma}^{\dagger}(\bm k)b_{\sigma}(\bm k) 
+ 
(\phi^\ast(\bm k)+3)b^\dag_{\sigma}(\bm k)
a_{\sigma}(\bm k)
\right]\,.
\end{eqnarray}

The Kubo formula for the conductivity is given by\cite{Paul03} 
\begin{equation}
\sigma_{xx}(\omega) = \frac {< j^D_x>}{iA_s(\omega + i0^+)}+
\frac {\Lambda_{xx}(\omega + i0^+)}{i\hbar A_s(\omega + i0^+)}\,,
\end{equation}
with $A_s=N_cA_c$ the area of the sample, and $A_c$ the area of the unit cell,
from which it follows that
\begin{equation}
\Re\sigma(\omega) = D\delta(\omega) + \frac {\Im \Lambda_{xx}(\omega + i0^+)}
{\hbar\omega A_s}\,,
\end{equation}
where $D$ is the charge stiffness which reads
\begin{equation}
D= -\pi \frac {<j^D_x>}{A_s} -\pi\frac {\Re \Lambda_{xx}(\omega + i0^+) }
{\hbar A_s}\,.
\end{equation}
The incoherent contribution to the conductivity $\Lambda_{xx}(\omega + i0^+)$ is obtained from $\Lambda_{xx}(i\omega_n)$,
with this latter quantity defined  as 
\begin{equation}
\Lambda_{xx}(i\omega_n) = \int_0^{\beta}d\,\tau e^{i\omega_n\tau}
<T_{\tau} j^P_{x}(\tau)j^P_x(0)>\,.
\end{equation}
The relevant quantity   $\Im\Lambda_{xx}(\omega + i0^+)$  is 
given by

\begin{eqnarray}
&&\Im\Lambda_{xx}(\omega + i0^+)=
\frac {t^2e^2a^2}{16\hbar^2}\int \frac {d\,\epsilon}{2\pi\hbar}\sum_{\bm k}
\sum_{\lambda_1,\lambda_2=\pm 1}
\nonumber\\ 
&&\left[n_F(\epsilon)-n_F(\epsilon+\omega\hbar)\right]
A^{\lambda_1}(\bm k,\omega+\epsilon/\hbar)
A^{\lambda_2}(\bm k,\epsilon/\hbar)
f(\bm k,\lambda_1,\lambda_2)\,,
\label{ImLambda}
\end{eqnarray}
with $A^{\lambda}(\bm k,\omega)$ given by
\begin{equation}
 A^{\lambda}(\bm k,\omega)=-2\Im G^{\lambda}_R(\bm k,\omega+i0^+)\,,
\end{equation}
 $G^{\lambda}(\bm k,i\omega)$ given by
\begin{equation}
 G^{\lambda}(\bm k,i\omega_n)=\frac {1}{i\omega_n-\Sigma(i\omega_n)
-(\lambda t \vert\phi(\bm k)\vert-\mu)/\hbar}\,,
\end{equation}
and $f(\bm k,\lambda_1,\lambda_2)$ given by
\begin{eqnarray}
f(\bm k,\lambda_1,\lambda_2)=2\vert \phi(\bm k)-3\vert^2
-\lambda_1\lambda_2\left[
(\phi^\ast(\bm k)-3)^2\frac {\phi^2(\bm k)}{\vert \phi(\bm k)\vert^2}
+
(\phi(\bm k)-3)^2\frac {(\phi^\ast(\bm k))^2}{\vert \phi(\bm k)\vert^2}
\right].
\end{eqnarray}
Using the fact that
\begin{eqnarray}
\sum_{\bm k}\phi(\bm k)g(\vert\phi(\bm k)\vert)=
\sum_{\bm k}\phi^\ast(\bm k)g(\vert\phi(\bm k)\vert)
=
\frac 1 3
\sum_{\bm k}\vert\phi(\bm k)\vert^2 g(\vert\phi(\bm k)\vert)
\end{eqnarray}
where $g(\vert\phi(\bm k)\vert)$ is an arbitrary function of the
absolute value of $\phi(\bm k)$, and the fact that in the
Dirac cone approximation one has
\begin{equation}
\frac {\phi^2(\bm k)}{\vert \phi(\bm k)\vert ^2}\simeq
e^{i2\pi/3}[\cos(2\theta)-i\sin(2\theta)]\,,
\end{equation}
and a similar expression for the complex conjugate expression 
$[\phi^\ast(\bm k)]^2/\vert \phi(\bm k)\vert ^2$, one obtains
a simplified expression for (\ref{ImLambda}), given by 

\begin{figure}
\begin{center}\includegraphics*[width=8cm]{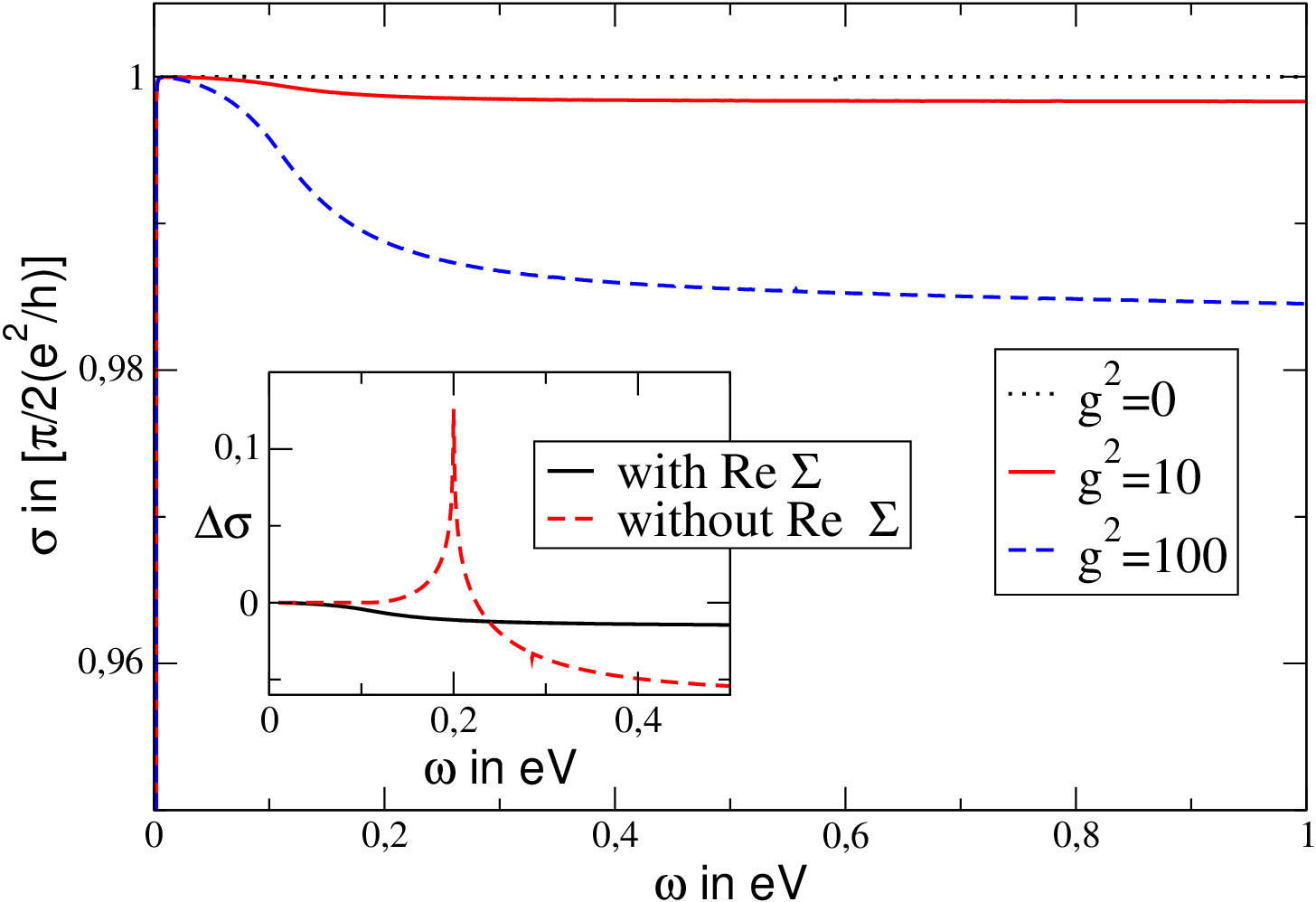}\end{center}
\caption{\label{cap:conductivity}(Color online) Optical 
conductivity for two different values of the coupling 
constant $g$ at zero doping ($T=\Gamma=10^{-4}$eV). 
Inset: The relative conductivity $\Delta\sigma=
\sigma(g=10)-\sigma(g=0)$ with and without $\Re \Sigma(\omega)$.}
\end{figure}

\begin{eqnarray}
&&\Im\Lambda_{xx}(\omega + i0^+)=
\frac {t^2e^2a^2}{16\hbar^2}\int \frac {d\,\epsilon}{2\pi\hbar}\sum_{\bm k}
\sum_{\lambda_1,\lambda_2=\pm 1}
\nonumber\\ 
&&\left[n_F(\epsilon)-n_F(\epsilon+\omega\hbar)\right]
A^{\lambda_1}(\bm k,\omega+\epsilon/\hbar)
A^{\lambda_2}(\bm k,\epsilon/\hbar)
\nonumber\\
&&[18-2\vert\phi(\bm k)\vert^2(\lambda_1\lambda_2+1)]\,.
\label{ImLambda_simp}
\end{eqnarray}

The expression (\ref{ImLambda_simp}) is valid only in the Dirac
cone approximation, and therefore one must replace
$t\vert\phi(\bm k)\vert$ by $v_F\hbar k$ and the integral
over the momentum can be easily done. If one ignores the effect
of phonons the calculation is straightforward, leading to
a conductivity of the form (at zero doping)\cite{Per07}
\begin{equation}
\Re\sigma(\omega) = \frac {\pi}2\frac {e^2}{h}
\left(
1-\frac {(\hbar \omega)^2}{18t^2}
\right)\tanh
\left(
\frac {\omega\hbar}{4k_BT}
\right)\,.
\label{free_sigma}
\end{equation}
One should note that from Eq. (\ref{free_sigma})
one has  $\Re\sigma(0)=0$ for finite $T$. This is not seen in
Fig. \ref{cap:conductivity} because the temperature
scale is very small.

The solution of Eq. (\ref{ImLambda_simp}) allows us to determine
the optical conductivity taking into account the effect of
Holstein phonons. In addition, we mimic the effect of impurities
by adding a small imaginary part $\Gamma$ to the self-energy
\begin{equation}
  \Im\Sigma(\omega) = \Im\Sigma_{ep}(\omega) - \Gamma\,,
\end{equation}
where $ \Im\Sigma_{ep}(\omega)$ is obtained from Eq.
(\ref{self_energy_ep}). 

There are two types of momentum integrals in (\ref{ImLambda_simp}),
which have the form
\begin{equation}
I_n(\epsilon/\hbar,\omega,\lambda_1,\lambda_2)=\int^{k_c}_0dk
\frac{k^{2n+1}}
{[(A-\lambda_1k)^2+B^2][(C-\lambda_2k)^2+D^2]}\,,
\end{equation}
with $n=0,1$. The analytical expressions are given in the appendix. 
The final energy integration can be done numerically.

In Fig. \ref{cap:conductivity}, the conductivity is shown for various
coupling constants $g$ at zero doping and low temperature $T$ and
damping $\Gamma$ due to impurity scattering.  There is a drop of the
conductivity (relative to the conductivity of a clean system) starting
at around the phonon energy $\omega_0$ and reaching a constant value
for $\omega\gapp2\omega_0$. Without the real part of the electronic
self-energy, there is a pronounced peak at twice the phonon
energy. This is shown in the inset of Fig. \ref{cap:conductivity},
where we plot $\Delta\sigma= \sigma(g=10)-\sigma(g=0)$ for
$T=\Gamma=0.0001$eV and including, respectively, not including 
$\Re \Sigma(\omega)$ in the above expressions. It is thus crucial to
include the full self energy in the renormalization of the particle
Green's function.

We thus obtain as main result that there remains no pronounced structure in the conductivity due to ZO-phonons if the full self-energy is used. We attribute this fact to the apparently asymmetric way the self-energy enters in the Green's function with respect to the electron (j=1) and hole (j=-1) channel which destroys possible interferences between the two carriers (see Eq. (\ref{self_energy_ep})).

\begin{figure}
\begin{center}\includegraphics*[width=8cm]{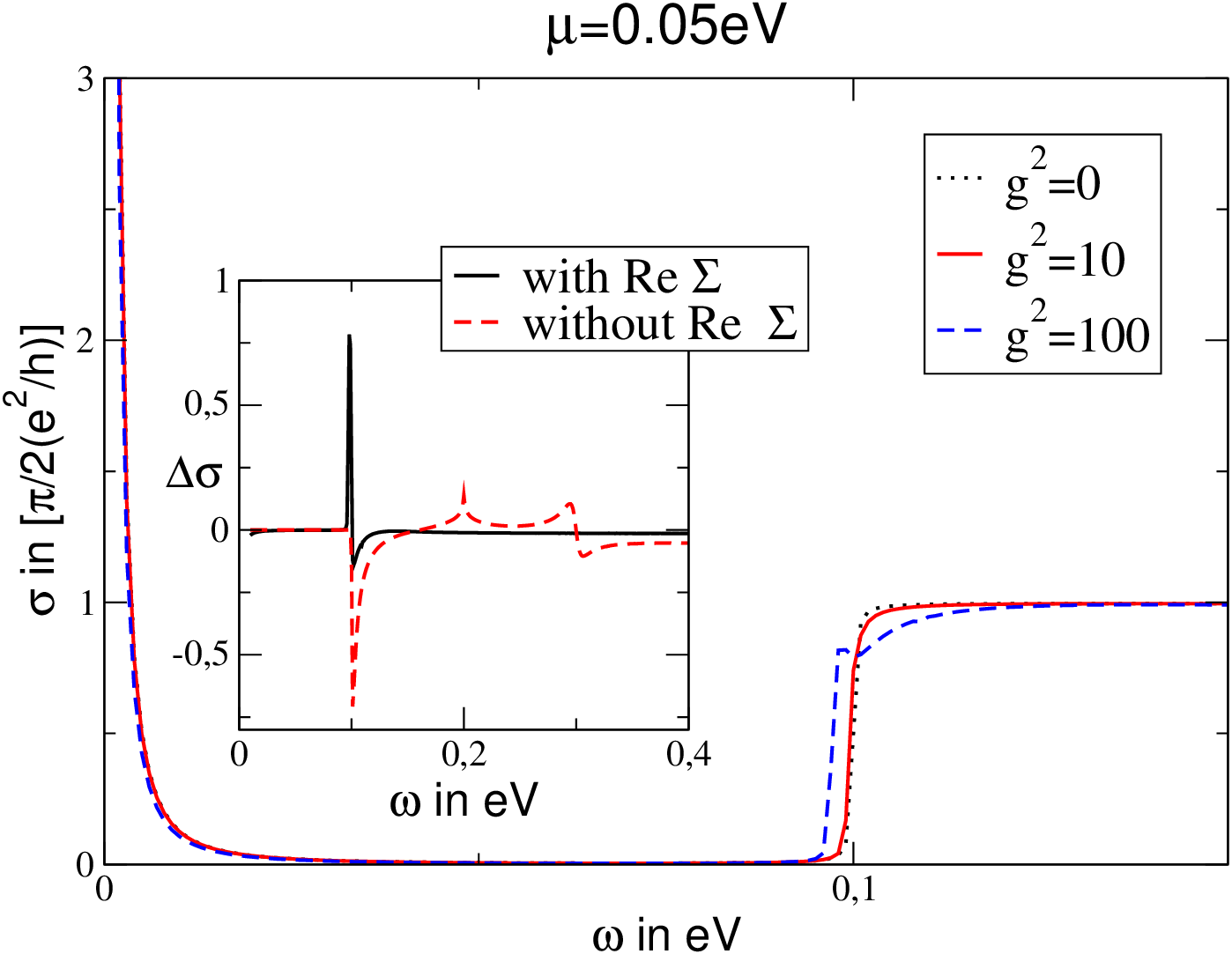}\end{center}
\caption{\label{cap:conductivity_mu0.05}(Color online) Optical 
conductivity for two different values of the coupling constant 
$g$ at $\mu=0.05$eV ($T=\Gamma=10^{-4}$eV. Inset: The 
relative conductivity $\Delta\sigma=\sigma(g=10)-\sigma(g=0)$ 
with and without $\Re \Sigma(\omega)$.}
\end{figure}

\begin{figure}
\begin{center}\includegraphics*[width=8cm]{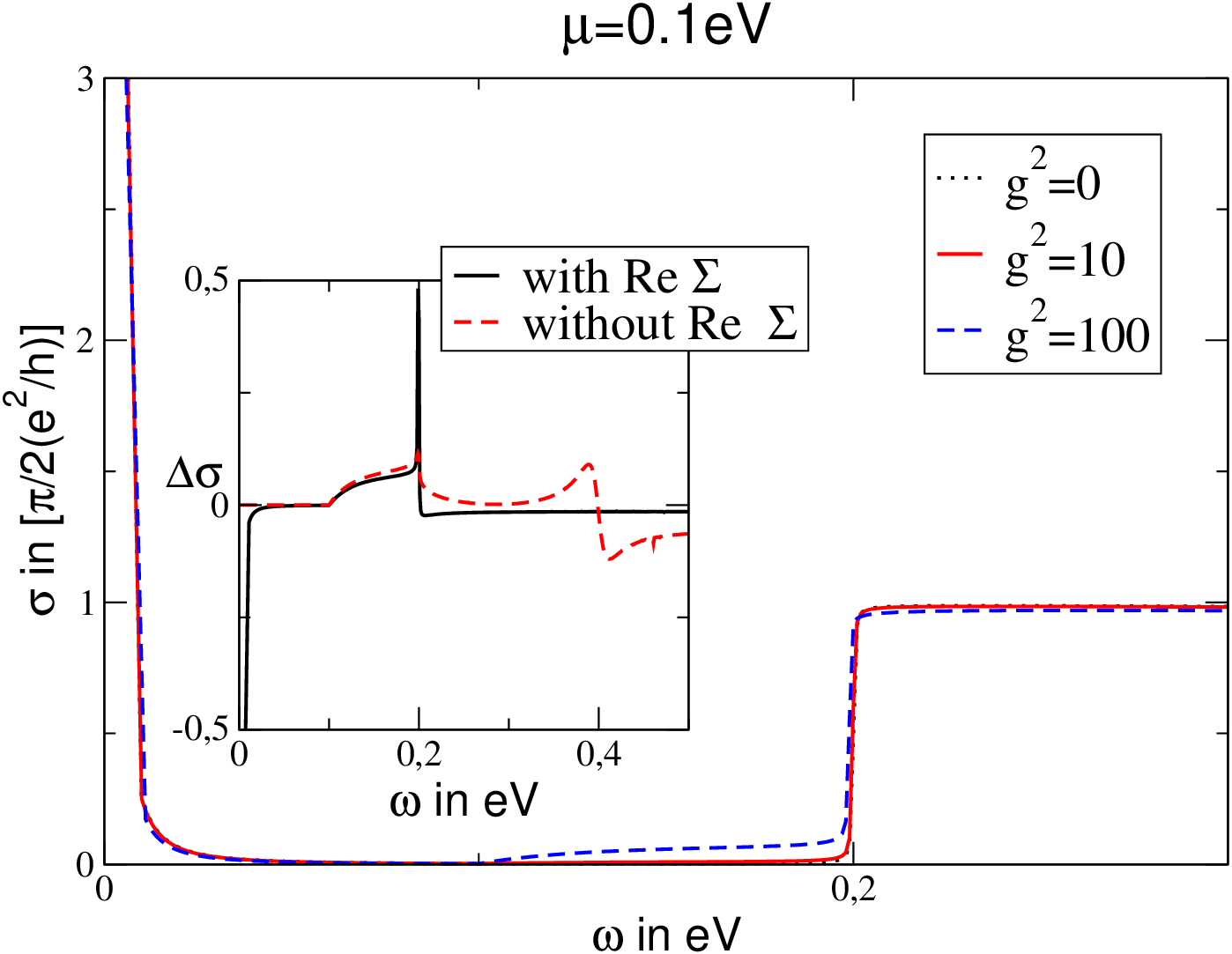}\end{center}
\caption{\label{cap:conductivity_mu0.1}(Color online) 
Optical conductivity for two different values of the 
coupling constant $g$ at $\mu=0.1$eV ($T=\Gamma=10^{-4}$eV. 
Inset: The relative conductivity $\Delta\sigma=
\sigma(g=10)-\sigma(g=0)$ with and without $\Re \Sigma(\omega)$.}
\end{figure}

In Figs. \ref{cap:conductivity_mu0.05} and \ref{cap:conductivity_mu0.1}, 
the optical conductivity is shown for different values 
of the coupling 
constant $g$ at $\mu=0.05$eV and $\mu=0.1$eV, respectively. The insets 
show the relative conductivity due to the electron-phonon 
interaction, $\Delta\sigma=\sigma(g=10)-\sigma(g=0)$, 
including, respectively, not including $\Re \Sigma(\omega)$ 
in the above expressions. Again we see a distinct difference 
in the result due to the renormalization of the electron-hole spectra.

It is clear that the results at finite chemical potential are markedly
different from the results at the neutrality point. At finite chemical
potential the system is characterized by a Drude like behavior
followed by a strong increase of the conductivity when the photon
frequency reach the value of twice the chemical potential. At zero
doping there is no Drude weight and the system response is
characterized only by inter-band transitions. We see only weak
renormalization of the Drude peak due to the Holstein phonons as well
as negligible effects at finite frequencies. We note that the results for
$\Re\sigma(\omega)$ when $g=0$ were first obtained by Peres {\it et
al.}\cite{Peres06} and by Gusynin {\it et al.}.\cite{Gusynin06}

\section{Summary}
In this paper, we have calculated the effect of Holstein polarons on the
electronic properties of graphene. Holstein polarons arise through the
coupling of out-of-plane optical (ZO) modes to the conduction
electrons, described as Dirac Fermions.  Throughout this work, we
assumed Migdal's theorem to be valid and calculated the electronic
self-energy within the first Born approximation. We find that the
Fermi velocity becomes renormalized within 1 \%.  

The main purpose of this work was to assess the effect of Holstein
phonons on the conductivity within the Kubo-formula. Due to the large
phonon-energy, electron scattering from Holstein phonons induces
interband transitions for usual carrier densities (corresponding to a
gate voltage of $\sim50$V) and can thus not be treating within the one-band
Boltzmann approach. We thus calculated the optical conductivity within
the Kubo-formula, employing the full Green's function but neglecting
vertex corrections. We find a pronounced kink-like peak at twice the
ZO-phonon energy if only the imaginary part of the self-energy is
considered. This peak vanishes when the real part of the self-energy
is included. Further, we see only weak renormalization of the Drude
peak due to the Holstein phonons as well as negligible effects at
finite frequencies. We conclude that scattering from Holstein-phonons
can be neglected in usual transport properties.

The effect of lattice vibrations on the electronic properties of
graphene is still not fully understood. Especially the coupling of
substrate phonons\cite{Fratini} or in-plane oscillations to the
conduction electrons is interesting. In the later case, this will lead
to a non-diagonal electronic self-energy due to the current-current
coupling.
\section*{Acknowledgments}
The authors want to thank F. Guinea and A. H. Castro Neto for useful
discussions. This work has been supported by Ministrio de Educaci\'on y Ciencia (Spain) through
Grant No. FIS2004-06490-C03-01, the Juan de la Cierva Programme, and 
by the European Union through contract 12881 (NEST).
N.~M.~R.~Peres thanks the European Science Foundation Programme INSTANS 2005-2010, and
Funda\c{c}\~ao para a Ci\^{e}ncia e a Tecnologia under the grant PTDC/FIS/64404/2006.
\appendix
\section{ Momentum integrals}
The momentum integrals have the form ($n=0,1$)
\begin{equation}
I_n=\int\frac{k^{2n+1}dk}
{[(A-k)^2+B^2][(C-k)^2+D^2]}\;.
\end{equation}
The general solution yields
\begin{eqnarray}
I_n&=&\frac{-1}{G}\Big[F_1^n\tan^{-1}\left(\frac{A-k}{B}\right)+F_2^n\tan^{-1}\left(\frac{C-k}{D}\right)\nonumber\\
&-&F_3^n\ln\left((A-k)^2+B^2\right)-F_4^n\ln\left((C-k)^2+D^2\right)\Big]
\end{eqnarray}
where the denominator reads
$
G=(B^2+(A-C)^2)^2
+2((A-C)^2-B^2)D^2+D^4
$
and the factors are given by
\begin{eqnarray}
F_1^0&=&4D(A^3-2A^2C-2B^2C+A(B^2+C^2+D^2))\;,\nonumber\\
F_2^0&=&4B((B^2+(A-C)^2)C+(-2A+C)D^2)\;,\nonumber\\
F_3^0&=&-F_4^0=2BD(-A^2-B^2+C^2+D^2)\nonumber
\end{eqnarray}
and 
\begin{eqnarray}
F_1^1&=&4D(A^5-2A^4C+2B^4C+A^3(2B^2+C^2+D^2)+AB^2(B^2-3(C^2+D^2)))\;,
\nonumber\\
F_2^1&=&4B((B^2+(A-C)^2)C^3+C(-3(A^2+B^2)+2C^2)D^2+(2A+C)D^4)\;,
\nonumber\\
F_3^1&=&2BD(A^4-4A^3C-4AB^2C+B^2(B^2-C^2-D^2)+A^2(2B^2+3(C^2+D^2)))\;,
\nonumber\\
F_4^1&=&2BD(C^2(3(A^2+B^2)-4AC+C^2)-(A^2+B^2+4AC-2C^2)D^2+D^4)\;.
\nonumber
\end{eqnarray}

\end{document}